\documentclass[conference]{IEEEtran}
\IEEEoverridecommandlockouts
\usepackage{cite}
\usepackage{verbatim}
\usepackage{amsmath,amssymb,amsfonts}
\usepackage{braket}
\usepackage{siunitx}
\usepackage{float}
\usepackage{graphicx, enumitem}
\usepackage[caption=false, font=footnotesize]{subfig}
\usepackage{textcomp}
\usepackage[table,dvipsnames]{xcolor}
\usepackage{tikz}
\usepackage{tabularx}
\usepackage{bm}
\usepackage{soul}
\usepackage{makecell}
\usepackage{booktabs}
\usepackage{multirow}
\usepackage{tikz}
\usepackage[none]{hyphenat}
\usepackage{orcidlink}
\usepackage{colortbl}

\definecolor{Gray}{gray}{0.85}
\definecolor{LightCyan}{rgb}{0.88,1,1}

\newcolumntype{L}[1]{>{\raggedright\let\newline\\\arraybackslash\hspace{0pt}}m{#1}}
\newcolumntype{C}[1]{>{\centering\let\newline\\\arraybackslash\hspace{0pt}}m{#1}}
\newcolumntype{R}[1]{>{\raggedleft\let\newline\\\arraybackslash\hspace{0pt}}m{#1}} 
\newcolumntype{a}{>{\columncolor{Gray}}c}
\newcolumntype{b}{>{\columncolor{white}}c}

\def\BibTeX{{\rm B\kern-.05em{\sc i\kern-.025em b}\kern-.08em
    T\kern-.1667em\lower.7ex\hbox{E}\kern-.125emX}}

\newcommand{\angs}{\text{\r{A}}}

\newcommand{\seclabel}[1]{Section~\ref{#1}}
\newcommand{\tablabel}[1]{Table~\ref{#1}}
\newcommand{\figlabel}[1]{Figure~\ref{#1}}

\begin{document}

\title{On the Transfer of Knowledge in \\ Quantum Algorithms}

\author{
	\IEEEauthorblockN{Esther Villar-Rodriguez\IEEEauthorrefmark{1}\IEEEauthorrefmark{2}\orcidlink{0000-0003-3343-3737},
	    Eneko Osaba\IEEEauthorrefmark{2}\orcidlink{0000-0001-7863-9910},
        Izaskun Oregi\IEEEauthorrefmark{2}\orcidlink{0000-0002-3950-1668}, \\
        Sebastián V. Romero\IEEEauthorrefmark{2}\IEEEauthorrefmark{3}\textsuperscript{\textsection}\orcidlink{0000-0002-4675-4452} and
        Julián Ferreiro-Vélez\IEEEauthorrefmark{2}\IEEEauthorrefmark{3}\IEEEauthorrefmark{5}\orcidlink{0000-0003-4864-7623}
		}
	\IEEEauthorblockA{\IEEEauthorrefmark{2}TECNALIA, Basque Research and Technology Alliance (BRTA), 48160 Derio, Bizkaia, Spain}
    \IEEEauthorblockA{\IEEEauthorrefmark{3}Department of Physical Chemistry, University of the Basque Country UPV/EHU, Apartado 644, 48080 Bilbao, Spain}
    \IEEEauthorblockA{\IEEEauthorrefmark{5}EHU Quantum Center, University of the Basque Country UPV/EHU, 48940 Leioa, Spain}
    
	\IEEEauthorblockA{\IEEEauthorrefmark{1}Corresponding author. Email: esther.villar@tecnalia.com}}
\maketitle

\begingroup\renewcommand\thefootnote{\textsection}
\footnotetext{Present Address for S. V. R.: Kipu Quantum GmbH, Greifswalderstrasse 212, 10405 Berlin, Germany.}
\endgroup

\IEEEpubidadjcol	

\begin{abstract}
Quantum computing is poised to transform computational paradigms across science and industry. As the field evolves, it can benefit from established classical methodologies, including promising paradigms such as Transfer of Knowledge (ToK). This work serves as a brief, self-contained reference for ToK, unifying its core principles under a single formal framework. We introduce a joint notation that consolidates and extends prior work in Transfer Learning and Transfer Optimization, bridging traditionally separate research lines and enabling a common language for knowledge reuse. Building on this foundation, we classify existing ToK strategies and principles into a structured taxonomy that helps researchers position their methods within a broader conceptual map. \textcolor{black}{We then extend key transfer protocols to quantum computing, introducing two novel use cases—reverse annealing and multitasking Quantum Approximate Optimization Algorithm (QAOA)—alongside a sequential Variational Quantum Eigensolver (VQE) approach that supports and validates prior findings.} These examples highlight ToK’s potential to improve performance and generalization in quantum algorithms. Finally, we outline challenges and opportunities for integrating ToK into quantum computing, emphasizing its role in reducing resource demands and accelerating problem-solving. This work lays the groundwork for future synergies between classical and quantum computing through a shared, transferable knowledge framework.
\end{abstract}


\section{Introduction}
\label{sec:Introduction}

Traditionally, in the field of Artificial Intelligence (AI), problems have been addressed individually using methods that do not rely on any preliminary knowledge. This approach has proven effective in many real-world scenarios \cite{davenport2018artificial}. However, the increasing complexity of contemporary problems, driven by digital transformation, has created an ideal environment for researchers to explore alternative resolution paradigms.

In this context, AI is now going beyond basic reasoning, striving for genuine learning in its quest to reach the next level: Strong AI \cite{ng2020strong}, an intelligence indistinguishable from the human mind. This human-level intelligence currently rests on two main pillars: (1) augmenting processing capacities based on human brain-inspired hardware and software, referred to as neuromorphic computing \cite{markovic2020physics}, to deliver skills closer to human cognition by emulating the nervous system’s functioning; and (2) inferring complex relational knowledge as well as accepting, as input, more enriched data representations such as ontologies or graphs. Nevertheless, implementing real intelligence demands progress not only in knowledge acquisition and methods to exploit it but also in knowledge retention to become efficient in information processing \cite{taherdoost2023artificial}.

This leads us inevitably to the reuse of knowledge: How much new information is \textit{truly} novel? Are machines creating knowledge or simply solving complex mathematical problems? This is where the discipline that leverages previous knowledge to accelerate and enhance the quality of new knowledge comes into play: Transfer Learning (TL, \cite{pan2009survey}) in the field of Machine Learning (ML) and its counterpart in Optimization, known as Transfer Optimization (TO, \cite{gupta2017insights}).

On this basis, and in both cases, the transfer of knowledge (ToK) is usually described as a mechanism to exploit the knowledge gained from completing previous tasks when facing new ones. However, this definition is not exhaustive. TL and TO also rely on multitasking schemes where all solvers are simultaneously running \cite{ong2016evolutionary,zhang2018overview}, deriving and sharing knowledge in parallel. Whatever the strategy, transfer is likely to become the cornerstone of more efficient processing.

The field of Quantum Computing (QC, \cite{gyongyosi2019survey}) is generating high expectations across both scientific and industrial communities. This emerging paradigm promises to fundamentally reshape how computing is conceived and executed. \textcolor{black}{In a nutshell, quantum computing leverages the dynamics of quantum systems to execute algorithms that exploit uniquely quantum-mechanical resources. In this paradigm, qubits serve as the fundamental building blocks of computation. A qubit is a mathematical abstraction of a quantum system and represents quantum properties such as superposition and entanglement.}

\textcolor{black}{Superposition reflects the probabilistic nature of quantum states, allowing a qubit to exist in a combination of the logical states 0 and 1 simultaneously, unlike classical bits, which are restricted to one state at a time. Entanglement, in turn, establishes non-classical correlations between qubits, creating a composite system that must be treated as a whole rather than as individual components.}

QC is still in the early stages of a gradual democratization process \cite{seskir2023democratization}. Currently, most available quantum devices fall under the Noisy Intermediate-Scale Quantum (NISQ) era \cite{preskill2018quantum}, a term coined to describe quantum processors characterized by a limited number of qubits, susceptibility to noise, and insufficient advancement for fault-tolerant computing. Although democratization in quantum technology is often assumed, as there are increasing activities geared towards providing access through the cloud and supporting this access with complementary training, the real democratization of reaching more stakeholders with devices freely available is halfway through. In fact, the cost of quantum hardware is currently far from insubstantial. Extensive experimentation or industrial deployments entail huge expenses. Here, efficiency becomes even more urgent~\cite{horowitz2019quantum}.

{\color{black} In the pursuit of such efficiency, QC can benefit from the long-standing experience of classical AI, which offers both technical maturity and valuable lessons \cite{villar2023hybrid}. This foundation enables the development of robust hybrid methods \cite{callison2022hybrid} and the adaptation of knowledge transfer techniques that have proven effective in classical domains \cite{transfer_2025}.}

In particular, the concept of TL in quantum ML has been explored by some references \cite{Zen_2020, schuman2023towards, Wang_2021, buonaiuto2024quantum,mari2020transfer}. Additionally, some concepts from TO have been investigated in solving various optimization problems, especially in contexts where the tasks to be solved appear sequentially \cite{qaoa_transfer,shaydulin2023parameter,transfer_parameters_quantum,nguyen2025cross,qaoa_transfer,bp_transfer,transfer_vqe_2025,vqe_transfer,laura_transfer}. Although several studies have explored ToK in quantum computing, many lack a consistent formalism or shared terminology \cite{transfer_parameters_quantum, transfer_vqe_2025}, often using alternative terms like “\textit{parameter recycling}” that obscure their link to the broader ToK framework. This terminological fragmentation underscores the need for a unified language to align classical and quantum perspectives and enable systematic comparison.

In response, this work offers a self-contained reference for both new QC practitioners and classical ToK researchers. We unify the foundational concepts of ToK under a single formal framework with consistent notation, bridging ML and optimization lines of research. This enables a principled classification of prior and future efforts across both paradigms, and lays the groundwork for synergy and real knowledge transfer between classical and quantum domains. The paper is structured around four core objectives:

\begin{itemize}
    \item[\textbf{O1}] {\color{black}Unify ToK foundations through a joint mathematical formulation that encompasses both classical ML and optimization domains, synthesizing and extending existing field-specific formulations of ToK \cite{weiss2016survey}.}
    \item[\textbf{O2}] {\color{black}Use this unified framework to systematically conceptualize and organize the core components of any ToK strategy—namely: \textbf{when to transfer}, i.e., the conditions under which knowledge transfer is appropriate, \textbf{what to transfer}, i.e., the transferable elements between tasks or domains, and \textbf{how to transfer}, i.e., the mechanisms or operational modalities by which knowledge is reused. Each of these dimensions is formally characterized using the joint notation, allowing a principled classification of existing research and serving as a self-contained reference for new researchers entering the field.}
    \item[\textbf{O3}] Demonstrate the applicability of the proposed framework to quantum computing through three representative use cases. \textcolor{black}{Specifically, we introduce two novel applications—reverse annealing and multitasking Quantum Approximate Optimization Algorithm (QAOA)—and revisit sequential Variational Quantum Eigensolver (VQE). These case studies collectively showcase the flexibility of the ToK paradigm and provide empirical evidence of its potential to enhance diverse quantum algorithms.}
    \item[\textbf{O4}] {\color{black}Highlight promising research directions through a structured analysis of challenges and opportunities for ToK in QC, starting from several potential scenarios where ToK might be valuable. This aims to encourage the QC community to explore and adopt ToK methodologies as a way to accelerate algorithmic development and leverage knowledge across tasks and domains.}
\end{itemize}

{\color{black} The research plan followed three main processes. To address objectives \textbf{O1} and \textbf{O2} (detailed in \seclabel{sotaTOK}), a systematic literature review on knowledge transfer in classical computing was conducted using databases like Scopus, Web of Science, IEEE Xplore, and preprint servers such as arXiv. To meet \textbf{O3}, three use cases were developed in \seclabel{sec:sotaQTOK} covering both quantum annealing and gate-based paradigms with diverse problem-solving strategies. Finally, objective \textbf{O4}, addressed in \seclabel{sec:challenges}, synthesizes previous findings, presenting application scenarios that examine what knowledge to transfer, in which point of the pipeline to transfer it, and how to do so effectively.}

\section{Fundamentals on Transfer of knowledge} \label{sotaTOK}

Knowledge transfer involves formulating principles to share knowledge by leveraging computed information for similar problems. These principles act as a bridge between what has already been learned or tested and what is currently {\color{black} being explored \cite{gupta2017insights}}. 

As mentioned in the introduction, two paradigms have principally arisen regarding the AI cores: TL in Machine Learning and TO in Optimization. Regardless of the discipline, three core aspects must be addressed in transfer learning research, as identified by Pan and Yang in \cite{pan2009survey}:%Regardless of the discipline, there are three main research questions to be answered \cite{pan2009survey}: 
\begin{enumerate} 
    %\item \textbf{When to transfer}, distinguishing situations that are advantageous from those that are counterproductive. 
    %\item \textbf{What knowledge} should be transferred between domains or tasks.
    %\item \textbf{How to transfer}, i.e., the strategy to develop ToK. 
    {\color{black} \item \textbf{When to transfer}, which involves distinguishing situations where transfer is beneficial from those where it may be counterproductive.}
    {\color{black} \item \textbf{What knowledge} should be transferred between domains or tasks.}
    {\color{black} \item \textbf{How to transfer}, referring to the strategies used to develop transfer of knowledge (ToK).}
\end{enumerate}

The following three subsections (\ref{sec:when}, \ref{sec:what}, and \ref{sec:how}), elaborate on the core principles related to these questions. {\color{black} Finally, Subsection \ref{sec:trends} outlines the main trends explored by the scientific community in both TL and TO, highlighting representative approaches and strategies in these fields}.

As a last mention in this introduction to fundamentals and as a preliminary for the subsections to come, let us define the two main concepts on which the theory of ToK is formulated in this work:

\begin{itemize}
    \item A \textit{task} refers to the specific objective or problem that needs to be solved and its characterization, i.e., the elements that are initialized or formulated explicitly for the objective.
    \item A \textit{domain} refers to the environment or delimited space in which the task is mapped and the solving procedure will evolve. 
\end{itemize}

Tables \ref{tab:transfer_learning} and \ref{tab:transfer_opt} define mathematically the components of \textit{domain} $\mathcal{D}$ and \textit{task} $\mathcal{T}$ for both TL and TO. Note that we use the same symbols for homologous concepts to homogenize and simplify the notation.

\begin{table}[h]
    \centering
    \caption{Notation - Transfer Learning}\label{tab:transfer_learning}
        \begin{tabularx}{\linewidth}{lX}
        \toprule
          \textbf{Given}: \\
          $x$ & Input sample (a single vector of features) \\ \midrule
          \multicolumn{2}{c}{\textbf{DOMAIN $\mathcal{D}$}:} \\
          $\mathcal{X}$ & Feature space, i.e. $n$-dimensions where your variables live\\ 
          $X$ & Input space, such that $X \subset \mathcal{X}$, i.e. the space delimited by the input samples $\{x_1,x_2...x_m\}$ of the dataset\\
          $P(X)$ & Marginal distribution of the input space $X$ \\ \midrule
          \multicolumn{2}{c}{\textbf{TASK $\mathcal{T}$}:} \\
          $\mathcal{Y}$ & Label space, i.e. the set of all possible values for the label\\
          $Y$ & Output space, i.e. the labels $\{y_1,y_2,\dots,y_m\}$ associated to each sample in $X=\{x_1,x_2,\dots,x_m\}$\\
          $f(\cdot;\theta)$ & Predictive function, where $\theta$ are the hyperparameters of $f$ to be learnt. Probabilistically, $P(\mathcal{Y}|X)$ in supervised settings. \\   \bottomrule
         \end{tabularx}
\end{table}

\begin{table}[h]
    \centering
    \caption{Notation - Transfer Optimization}\label{tab:transfer_opt}
        \begin{tabularx}{\linewidth}{lX}
        \toprule
          \textbf{Given}: \\
          $x$ & Individual (a single vector of optimization variables)\\ \midrule
          \multicolumn{2}{c}{\textbf{DOMAIN $\mathcal{D}$}: } \\
          $\mathcal{X}$ & Search space, i.e. the vocabulary, the set of all available solutions through codification\\
          $X$ & Explorable search space, i.e. the search space limited by algorithm operation such that $X \subset \mathcal{X}$ \\ \midrule
          \multicolumn{2}{c}{\textbf{TASK $\mathcal{T}$}: } \\
          $X^*$ &  Optimal solutions such that $X^*=\arg \max f(x),\text{ }\forall x\in X$ \\
          $X^f$ & Feasible region such that $X_f=\{x \in X \,|\, g_k(x)\le 0, \text{ }\forall k\in\{1,\dots,|\vec{g}|\}, \text{ } h_l(x) = 0, \text{ } \forall l\in\{1,\dots,|\vec{h}|\}\}$ \\ \bottomrule
         \end{tabularx}
\end{table}

\subsection{When to transfer} \label{sec:when} %% reasons to resort to Transfer of knowledge

Let us assume that we had built a deep learning model to classify people as either joyful or sad, and we were considering adding new mood categories. It would be perfectly sensible to exploit the knowledge extracted in the initial run. In fact, if we examined the first filters of the neural network, we would discover patterns that define postural deviations in facial components. This represents cross-domain knowledge in computer vision~\cite{zhu2014weakly}.

Multiple examples of TL have been documented in the literature, particularly in the fields of Image Recognition and Natural Language Processing \cite{iqbal2017cross,regin2024fine,sulaiman2024evaluation}. However, not all similarities among tasks and domains are immediately apparent. Some similarities may be subtle and require preprocessing to become evident.

{\color{black} Therefore, it becomes sensible to devote time to discovering similarities when we are aware—or have the intuition, often drawn from experience—that a task may reappear in the future in a similar form with slight variations. Likewise, when there is a core of knowledge that remains valid across different use cases, considering ToK is a reasonable approach.}
%\textit{\textbf{When should we devote time to discovering similarities?}} When aware or having the intuition (drawn from experience) that a task will appear in the future in a similar form, with slight variations, or when there is a core of knowledge that remains valid across use cases, it is sensible to consider ToK. 
Violating this assumption, i.e., transferring knowledge when neither the domain nor the task is similar, will likely lead to encountering \textit{negative transfer} between the source and target tasks (\textit{origin} and \textit{destination} of the transfer) \cite{zhang2022survey,kumar2024enhancing,ahn2024reset}. As stated in \cite{pan2009survey}: \begin{quote}
\textit{``When the source domain and target domain are not related to each other, brute-force transfer may be unsuccessful. In the worst case, it may even hurt the performance of learning in the target domain, a situation which is often referred to as negative transfer.''}
\end{quote}

Bearing this in mind, the ToK may assist, particularly:
\begin{itemize} 
    \item \textbf{When} facing very complex search spaces in optimization problems \cite{osaba2022evolutionary}, 
    \item \textbf{When} there is a risk of overfitting, the model generalizes poorly, and the data set is relatively small in ML tasks (typically due to annotation expense or privacy concerns) \cite{pan2009survey}, and 
    \item \textbf{When} speed in processing is a must-have \cite{tan2018survey}. 
\end{itemize}

Having said this, literacy in ToK has postulated some conditions that, in case of being met, turn the problem to be solved into a good candidate for knowledge transfer.

In ML, and given the definitions provided in Table \ref{tab:transfer_learning}, \textit{when} is proposed under this classification \cite{pan2009survey}:
\begin{itemize}
    \item \textbf{Inductive transfer learning}: Given a source domain $\mathcal{D}_S$ and a corresponding learning task $\mathcal{T}_S$, a target domain $\mathcal{D}_T$ and a learning task $\mathcal{T}_T$, inductive transfer learning aims to improve the learning of the target predictive function $f(\cdot;\theta)_T$ in $\mathcal{D}_T$ using the knowledge of $\mathcal{D}_S$ and $\mathcal{T}_S$ given that $\mathcal{T}_S \neq \mathcal{T}_T$. A typical example of inductive transfer is sharing the feature representation layers in a neural network for two tasks in a multitask setting: a regression and a classification. This involves one backbone with two prediction heads.
    
    \item \textbf{Transductive transfer learning}: Given a source domain $\mathcal{D}_S$ and a corresponding learning task $\mathcal{T}_S$, and a target domain $\mathcal{D}_T$ and a corresponding learning task $\mathcal{T}_T$, transductive transfer learning aims to improve the learning of the target predictive function $f(\cdot;\theta)_T$ in $\mathcal{D}_T$ using the knowledge from $\mathcal{D}_S$ and $\mathcal{T}_S$, provided that $\mathcal{T}_S = \mathcal{T}_T$ and $\mathcal{D}_S \neq \mathcal{D}_T$. Transductive learning is most commonly exemplified in situations where there is a large amount of unlabeled data in $\mathcal{D_T}$, which hinders the training of $f(\cdot;\theta)_T$ with data from $\mathcal{D_T}$ and also the retraining of a pre-trained predictor $f(\cdot;\theta)_S$ to adjust to the new domain $\mathcal{D_T}$. To circumvent this situation, \textit{domain adaptation} (DA) techniques are commonly applied to achieve a domain translation, i.e., a meaningful correspondence or alignment, between $\mathcal{D_S}$ and $\mathcal{D_T}$. 

    \item \textbf{Unsupervised transfer learning}: Given a source domain  $\mathcal{D}_S$ with a learning task $\mathcal{T}_S$, a target domain  $\mathcal{D}_T$ and a corresponding learning task $\mathcal{T}_T$, unsupervised transfer learning aims to help improve the learning of the target predictive function in $\mathcal{D}_T$ using the knowledge in $\mathcal{D}_S$ and $\mathcal{T}_S$, where $\mathcal{T}_S \neq \mathcal{T}_T$ and $Y_S$ and $Y_T$ are not available. For instance, assuming that somehow datasets from $\mathcal{D}_S$ and $\mathcal{D}_T$ share a common underlying structure that can be leveraged to perform transfer learning in a clustering or dimensionality reduction task. 
\end{itemize} 

In optimization, analogously to transfer in ML, we should meet the following conditions:
\begin{itemize}
    \item \textbf{Overlapping in domains}: $\mathcal{D}_S \sim \mathcal{D}_T$, %by $\mathcal{X}_S \cap \mathcal{X}_T  \neq \varnothing$, 
    i.e. common codification scheme, and ${X}_S \cap {X}_T  \neq \varnothing$, i.e. a shared explorable search space.
    \item \textbf{Overlapping in tasks}: $\mathcal{T}_S \sim \mathcal{T}_T$, by means of $X^f_T \cap X^f_S \neq \varnothing$ and ideally  $X^*_T \cap X^*_S \neq \varnothing$, i.e. source and target task share regions of feasible and optimal space. 
\end{itemize}
where the more overlapping, the more contribution of ToK is expected \cite{ong2016evolutionary}.

\begin{table*}[h!]
\centering
\caption{Approaches for transfer of knowledge.}\label{Tab:Twhat2}
\def\arraystretch{1.25}
\begin{tabularx}{.9\linewidth}{ccc}
    %\arrayrulecolor{Bittersweet}
    \toprule \multicolumn{2}{c}{\textbf{WHAT TO TRANSFER}} & \multirow{2}*{\textbf{BRIEF EXPLANATION}} \\
     \multicolumn{1}{c}{\textbf{ML}} & \multicolumn{1}{c}{\textbf{Optimization}} &  \\
     \toprule
%     \arrayrulecolor{Black}
     \multicolumn{1}{a|}{Instance} & \multicolumn{1}{a|}{Individual} & \multicolumn{1}{l}{Sharing samples or individuals} \\ 
     %\hline
     \midrule
     \multicolumn{1}{a|}{Feature representation} & \multicolumn{1}{a|}{Search representation}  &  \multicolumn{1}{l}{Transferring a ``good'' learnt representation or aligning search space representations} \\
     \midrule
     %\hline
     \multicolumn{2}{a|}{Parameter} &  \multicolumn{1}{l}{Discovering shared (hyper-)parameters or operators} \\ %% eneko, esto se podría? se hace?
     %\hline
     \midrule
     \multicolumn{2}{a|}{Relational knowledge} & \multicolumn{1}{l}{Linking data or mapping relationships} \\
     %\arrayrulecolor{Bittersweet}
     \bottomrule
\end{tabularx}
\end{table*}

\subsection{What to transfer}\label{sec:what}
The transversal knowledge, the \textit{gold}, that boosts the learning or solving procedure and improves performance {\color{black} is outlined in \tablabel{Tab:Twhat2} (which builds upon the classification proposed in \cite{pan2009survey}, extending it to include both machine learning and optimization domains):} %(which is an adaptation of the classification proposed in \cite{pan2009survey} to encompass both ML and optimization areas):

\begin{itemize}
    \item \textbf{Instance or individual transfer}: Bringing samples or individuals (i.e. a subset of $X$) from $\mathcal{D}_S$ with some adjustment in the transfer, via re-weighting or importance sampling in ML, or mapping the encodings in optimization.
    \item \textbf{Feature or search representation transfer}: Inferring a (sometimes low-dimensional) common feature representation to be shared across tasks in ML or a common search space representation in optimization, mainly to accommodate multitasking schemes. This involves transferring $\mathcal{X}$ in both cases.
    \item \textbf{Parameter transfer}: Assigning the hyperparameters $\theta$ in $f_S(\cdot;\theta)$ to the model $f_T(\cdot;\theta)$ in ML or helping set the prior distributions of them in both ML and optimization algorithms.
    \item \textbf{Relational knowledge}. The motivation lies in the fact that there are relationships among the data $X_S$ that can be exploited in $\mathcal{T}_T$ as the relational structure holds or there exists a common sub-structure \cite{kumaraswamy2015transfer}.
\end{itemize}

\subsection{How to transfer} \label{sec:how}
Once the rationale and main concepts are defined, it is time to design the strategy for knowledge sharing. In this regard, two main conceptualizations can be distinguished:
\begin{itemize}
    \item \textbf{Sequential Transfer}: Task $\mathcal{T}_S$ precedes task $\mathcal{T}_T$, meaning that the knowledge created in $\mathcal{T}_S$ is subsequently imported into $\mathcal{T}_T$.
    
    \item \textbf{Multitasking}: Knowledge is created and shared during the parallel and simultaneous solving of a set of tasks. Therefore, multitasking relies on omnidirectional transfer, with tasks interchangeably playing the roles of source and target. In the field of optimization, when the pool of tasks represents different formulations of the same problem, this approach is known as \textit{Multiform Optimization} \cite{feng2024review}.
\end{itemize}

\subsection{Developments in Classical Computation} \label{sec:trends}
\begin{figure}[h!]
    \centering
    \includegraphics[width=\linewidth]{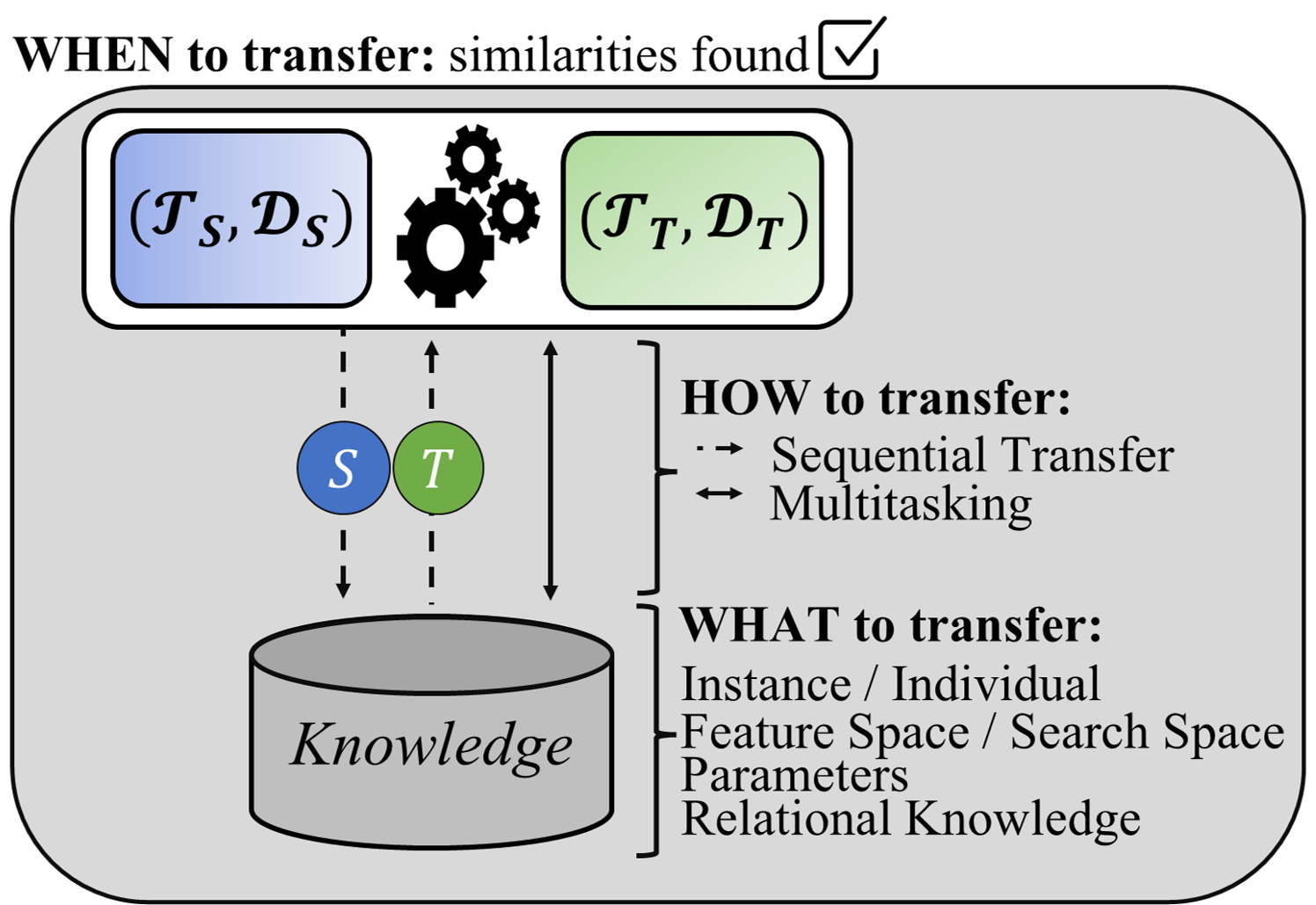}
    \caption{Transfer of knowledge philosophy.}
    \label{fig:tlp}
\end{figure}

Once explained the main concepts of ToK, graphically represented in Fig. \ref{fig:tlp}, the following paragraphs will briefly outline the trends followed by the scientific community in both TL and TO fields. In this context, many variants can be crafted by blending the previously mentioned approaches and strategies.

Regarding TL, it is commonly applied to deep learning (DL) algorithms, where deep neural networks are particularly popular. Training a model from end to end and from scratch is usually expensive, as it requires large labeled datasets and extensive computational resources. To reduce this cost, many TL approaches have been developed in recent years \cite{gholizade2025review, zhuang2020comprehensive, yan2024comprehensive}. Among these, instance and parameter-based frameworks are the most studied techniques. In terms of \textbf{instance transfer} approaches, DA \cite{csurka2017comprehensive,liu2025interpretable,li2024comprehensive} is one of the top applications. In general, DA is a transductive method that aims to exploit labeled data in the source domain $\mathcal{D}_S$ to learn a classifier $f(\cdot;\theta)_T$ for unlabeled instances in the target domain $\mathcal{D}_T$. 
    
While \textbf{sequential transfer} seems the most natural approach to address (DA) problems, \textbf{multitask transfer} is being employed to mitigate performance decay due to domain shift \cite{Elsahar_2019}, a phenomenon that occurs when the joint probability distributions of observed data and labels $P(\mathcal{Y}|X)$ differ between $\mathcal{D}_S$ and $\mathcal{D}_T$. %This scenario is, therefore, halfway between transductive and inductive transfer learning. 
Particularly, multi-source DA involves data collected from multiple sources with different distributions $P(X)$, which are aligned through \textit{latent space alignment} or \textit{intermediate domain generation} so that the learned $f(\cdot;\theta)_S$ can be better transferred to $f(\cdot;\theta)_T$ \cite{sun2015survey,zhao2020multi,zhang2024multi}. Conversely, data in $\mathcal{D}_S$ could be synthetically generated.

\textit{Adversarial transfer learning} focuses on using deceptive inputs to learn a more robust common representation, thereby boosting predictive performance on $\mathcal{T}_T$ \cite{deng2021adversarial,yu2019transfer,tang2024wind}. Meanwhile, \textbf{parameter transfer} approaches have been increasingly adopted in DL for inductive transfer \cite{cao2018preprocessing, wu2018convolution, liu2019convolutional}. In these approaches, the parameters of the initial layers of a pre-trained deep neural network are shared as a generic \textbf{feature representation}, and fine-tuning is performed on the remaining layers in the target network in a \textbf{sequential transfer} manner.

Another well-known field in which TL has shown its potential is \textit{reinforcement learning} \cite{zhu2023transfer,zamfirache2023q,kadamala2024enhancing}. Similar to instance transfer, sharing expert experiences or trajectories in a teacher-student approach via imitation learning \cite{hussein2017imitation} is probably the most familiar method, especially when efficient exploration is needed and $\mathcal{D}_S=\mathcal{D}_T$. Alternatively, \textit{policy distillation} \cite{zhu2023transfer} involves transferring one or more action policies to an untrained network with the objective of mimicking the teacher’s behavior. \textbf{Multitask} policy distillation has proven efficient in consolidating multiple task-specific policies into a single policy through replay memory buffers and evenly interleaved training across all tasks~\cite{rusu2015policy}.      
    
In reference to TO algorithms, most works propose \textbf{individual transfer} in a multitask or sequential setting. Regarding \textbf{multitasking}, \textit{multipopulation-based} \cite{song2019multitasking, cheng2017coevolutionary} and \textit{multifactorial} \cite{chen2024multi,li2024multifactorial} methods are the most paradigmatic. In multipopulation-based approaches, each population is assigned to a specific task $\mathcal{T}$, which evolves independently, allowing for individual migrations every $N$ iterations. Conversely, multifactorial methods rely on a joint search process involving a diverse population of individuals, each skilled at a specific task $\mathcal{T}$. These individuals are combined through the crossover operator to produce crossbred individuals, facilitating the ToK.

In these frameworks, complete domain overlap ($\mathcal{D}_S = \mathcal{D}_T$) provides the shortest path to transfer. However, as suggested by multiform proposals \cite{jiao2022multiform}, partial domain overlap when exploring the same problem search space via different sets of operators ($X_S \neq X_T$) may be more effective for exceptionally complex tasks \cite{jiao2022multiform, wu2022evolutionary}. Regardless of the trend, it is essential to maintain at least a common vocabulary when $\mathcal{X}_S \neq \mathcal{X}_T$ by creating a common \textbf{search space representation} as a shortcut to $\mathcal{D}_S=\mathcal{D}_T$. Since multitasking is prone to negative transfer, significant research has been conducted to mitigate this issue, such as mechanisms that automatically regulate the flow of information between different tasks~\cite{bali2019multifactorial, osaba2022evolutionary}.
    
With respect to \textbf{sequential} techniques, individuals to transfer are commonly chosen in an elitist manner, influencing the search in $\mathcal{T}_T$ by leveraging good solutions from $\mathcal{T}_S$ or, if found, $X^*_S$. These solutions are used to \textit{seed the initial population} \cite{xue2021evolutionary} or to stimulate an evolving population through \textit{dynamic injection} performed periodically \cite{louis2004learning}. Similar to the unified search space in multitasking, sequential transfer requires a mapping strategy to project solutions or preliminary knowledge that best matches the problem at hand from $\mathcal{D}_S$ to $\mathcal{D}_T$~\cite{qi2022evolutionary, feng2017autoencoding}.

\begin{figure*}[!h]
 \centering
 \includegraphics[width=0.93\linewidth]{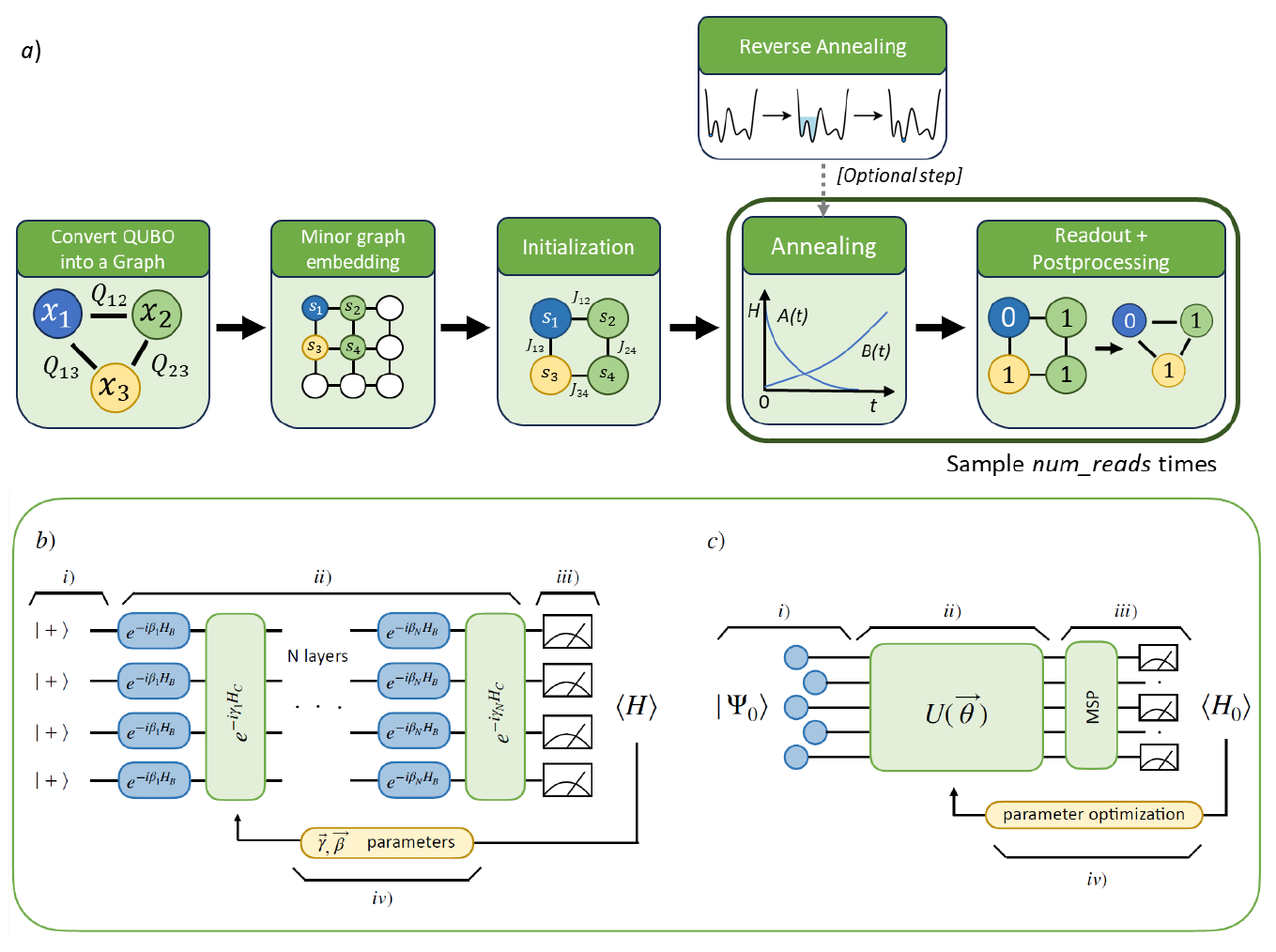}
 \caption{A schematic overview of the quantum algorithms employed. (a) Quantum annealing. Figures (b)-(c) illustrate the architecture of the variational algorithms utilized. Both algorithms can be systematically divided into the following stages: i) state initialization, ii) state manipulation, iii) measurement, and iv) a classical optimization subroutine. (b) The QAOA protocol applies a sequence of parameterized layers, alternating between the driver Hamiltonian $H_B$ and the problem Hamiltonian $H_C$, to an initial mixed state. (c) The VQE algorithm evolves an arbitrary initial state $|\Psi_0\rangle$ through a parameterized ansatz $\mathcal{U}(\theta)$, tailored to approximate the target ground state.}
 \label{fig:schemes}
\end{figure*}

\section{Quantum algorithms: Discovering potential synergies} \label{sec:sotaQTOK}

At the present time, there are two main accessible quantum computing paradigms: \textbf{quantum annealing} (QA) and \textbf{universal gate-based} quantum computing. 

First proposed in 1988, QA \cite{DeFalco_1988, APOLLONI1989233, FINNILA1994343, Kadowaki_1998} encodes problems using the Ising model, whose corresponding Hamiltonian can be used to map optimization problems or obtain lowest-energy samples \cite{Lucas_2014, moll2018quantum, 3dbpp2023, romero2025bias}. Depicted in \figlabel{fig:schemes}.a, the QA process begins by representing the problem as an energy landscape. Initially, the quantum system is prepared in a superposition of all possible solutions, representing a high-energy state. As the system evolves, the Hamiltonian is adjusted, allowing the system to explore the energy landscape. The goal is to guide the system towards the minimum energy state, which corresponds to the optimal solution of the problem.
Advances in quantum technologies have contributed to the construction of intermediate-scale Quantum Annealers for programmable use. In this context, the Canadian company D-Wave Systems has developed commercial devices based on QA, being the main provider of this technology. To date, these systems are the most widely used for optimization problems approached from a quantum perspective, with interesting research published in fields such as finance \cite{mugel2022dynamic}, logistics \cite{osaba2024solving}, and industry \cite{carugno2022evaluating}. Currently, the most advanced computer from D-Wave Systems, called \textit{Advantage\_system6.4}, consists of 5,612 superconducting qubits arranged in a Pegasus topology \cite{boothby2020next}. However, the launch of a new system, named \textit{Advantage2}, which will feature 7,430 qubits arranged in a Zephyr topology is scheduled for late 2025 \cite{Advantadge2}.

In contrast to QA, gate-based quantum computing, inspired by classical logic gates, operates by driving the quantum system within Hilbert space through a finite sequence of unitary gates, enabling universal computation. This approach requires precise control over quantum states, necessitating high-fidelity gates to implement complex and deep algorithms effectively. However, current hardware limitations restrict the range of feasible algorithms, making fault-tolerant algorithms like Shor's and Grover's, which demand high-precision control, currently impractical. 

To address these challenges, variational quantum algorithms (VQAs) have gained popularity in recent years \cite{Cerezo_2021}. These hybrid algorithms combine classical and quantum resources to exploit the quantum Hilbert space by parameterizing quantum gates. The parameters are optimized classically by minimizing a predefined cost function. By incorporating classical optimization into the process, hybrid algorithms enhance the robustness of the protocol. The classical optimization absorbs imperfections in the minimization process, making these algorithms more resilient to hardware-induced errors. Concrete examples of VQA are the QAOA \cite{Farhi2014} and the VQE \cite{Peruzzo2014}. The former was introduced to return approximate solutions for combinatorial optimization problems, whereas the latter was originally conceived to find the ground state energy of a given Hamiltonian, a central problem in quantum chemistry. In a more technical way:
\begin{itemize}
    \item \textit{QAOA} \cite{Farhi2014} aims to solve combinatorial optimization problems encoded in the ground state of the Hamiltonian $H_C$.  Starting from the ground state $| \Psi_0\rangle$ of the mixing Hamiltonian $H_B$, a parameterized circuit is applied to drive the system towards the target state $| \Psi_t(\vec{\gamma},\vec{\beta})\rangle = U_B(\gamma_p)U_C(\beta_p)...U_B(\gamma_0)U_C(\beta_0) |\Psi_0\rangle$, where $p$ denotes the algorithm's depth (see Fig. \ref{fig:schemes}.b). Parameters $\beta=\{\beta_i\}_{i=1}^p$ and $\gamma=\{\gamma_i\}_{i=1}^p$ are tuned using classical optimization subroutines, maximizing the cost function $\braket{H_C}_{\vec{\beta},\vec{\gamma}}=\braket{\Psi(\vec{\beta},\vec{\gamma})|H_C|\Psi(\vec{\beta},\vec{\gamma})}$.
    
    \item \textit{VQE} \cite{Peruzzo2014}, aims to find the Hamiltonian ground state by driving the system from an easy-to-prepare initial state $|\Psi_0\rangle$ to the target state $\ket{\Psi_T(\vec{\theta})}=U(\vec{\theta})\ket{\Psi_0}$, where $U(\vec{\theta})$ is the problem-specific ansatz  (see Fig. \ref{fig:schemes}.c). The parameters  $\theta=\{\theta_i\}_{i=1}^N$ are optimized by minimizing the energy expectation value of the Hamiltonian, defined as $\braket{H}_{\theta}=\braket{\Psi(\theta)|H|\Psi(\theta)}$, thought classical optimization techniques.
\end{itemize}
%\figlabel{fig:qaoa} and \ref{fig:vqe} describe the generic workflow of both methods.

\begin{table*}[!t]
	\centering
	\caption{Main characteristics of each UC and results. Results are depicted using the binomial composed of the source of knowledge (with \texttt{--} in the absence of knowledge transfer) and the results obtained. IQR: Interquartile Range, ICA: Iteration for Chemical Accuracy.}\label{tab:exp}
	\resizebox{1.92\columnwidth}{!}{
            \begin{tabularx}{\linewidth}{lcccc>{\centering\arraybackslash}Xcc}
			\toprule[1.5pt]
                & {\textbf{Solving Scheme}} & {\textbf{What}} & {\textbf{How}} & {\textbf{Point of Transfer}} & {\textbf{Parameterization}} & \multicolumn{2}{c}{\textbf{Performance analysis}}\\\midrule
                & & & & & & Source of Knowledge & (median, IQR) \\
                \cmidrule{7-8}

                \textbf{UC1} & QA & Individual & Sequential & Annealing & \makecell{\textit{\# Reads}: 1000 \\ \textit{Hold Time}: 100 \\
                \textit{s\_target}: 0.5\\ \textit{Ramp slope}: 2.00 \\ \textit{Reinitialize}: True} & \makecell{\texttt{--} \\ \textbf{\texttt{MaxCut\_50\_7}}  \\
                \texttt{MaxCut\_50\_9} \\ \texttt{MaxCut\_50\_15}  \\ \texttt{MaxCut\_50\_25} \\ \texttt{MaxCut\_50\_50} \\ {\color{black} \texttt{MaxCut\_50\_100}}}   & \makecell{(581, 10.0) \\ \textbf{(601, 1.5)}  \\
                (599, 0.0) \\ (590, 2.0)  \\ (588, 1.0) \\ (586, 1.75) \\ {\color{black} (577, 9.25)}}\\ \midrule

                & & & & & & Source of Knowledge & success prob. \\
                \cmidrule{7-8}
                \textbf{UC2} & QAOA & Parameters & Multitasking & State manipulation & \makecell{\textit{\# layers}: 2 \\ \textit{\# transfers}: 4 \\ 
                \textit{steps per transfer\_layer}: 20}  & \makecell{\texttt{--} \\ \textbf{Transfer\_static\_QAOA} \\Transfer\_evol\_QAOA} & \makecell{32.48 $\pm$ 0.59 \\ 34.71 $\pm$ 0.50 \\ 33.23 $\pm$ 0.95}\\ \midrule
                
                & & & & & & Source of Knowledge & ICA \\
                \cmidrule{7-8}
                \textbf{UC3} & VQE & Parameters & Sequential & State Initialization & \makecell{\textit{\# Reads: 100} \\ \textit{Optimizer}: SPSA \\ \textit{Init. state}: Hartree-Fock \\\textit{Ansatz}: UCC} & \makecell{\texttt{--} \\ \makecell{Optimal parameters\\of previous step}} & \makecell{29 \\ 2} \\\bottomrule[1.25pt]
		\end{tabularx}
    }
\end{table*}

\subsection{What to start with? Preliminary experimentation}

As part of this research, \textcolor{black}{we conducted three use cases (UCs)} to demonstrate the potential of ToK in the field of QC. \textcolor{black}{We selected these UCs} from the two primary quantum computing paradigms: gate-based and annealing-based computing. The first UC (UC1) involves the design of a sequential-individual transfer protocol within an annealing framework, highlighting when and how such transfer can be advantageous. The following two use cases examine transfer protocols in the main variational quantum computing paradigms: the VQE and QAOA. In the second use case (UC2), we propose an individual-based multitasking protocol for QAOA, investigating whether parallel transfer across similar graph instances can enhance optimization performance. Finally, to ensure a comprehensive analysis, the third UC (UC3) investigates a parameter-based sequential transfer strategy within the VQE algorithm, in line with recent studies in the literature \cite{vqe_transfer}. In this context, we explore the impact of sequential transfer techniques in molecular quantum computing, extending previous approaches by incorporating an experimental evaluation on real quantum hardware.
%Using an IBM quantum device, we evaluate the benefits of parameter transfer in the context of a chemistry simulation.}

To facilitate reproducibility, \tablabel{tab:exp} summarizes the description of the three UC, highlighting their principal characteristics: the solving scheme as well as the \textit{what} and \textit{how} to transfer. Furthermore, the parameterization and the results obtained are also described. We now dive into each UC for the sake of completeness.

\textbf{UC1: QA-Reverse Annealing, Individual-based, Sequential Transfer.} Framed in the QA paradigm, UC1 explores the efficiency of the reverse annealing (RA, \cite{Reverse}) for conducting sequential transfer of knowledge.

In a nutshell, RA is a method that allows the annealing process to go backward from a given state before moving forward to a new state. This involves performing a local search to refine a classical state. For individual-based transfer of knowledge to be utilized via RA, this classical state must be a solution obtained from a prior solver. Specifically, the main objective of this UC is to analyze the contribution and sensitivity of feeding RA with results obtained from solving similar problems.

To carry out UC1, we have utilized the well-known Maximum Cut problem (MCP, \cite{bodlaender2000complexity}), a problem widely employed in QC because of its simplicity. Specifically, the MCP is a graph partitioning problem whose objective is to find a cut such that the number of edges lying between the two subsets is maximized. In this case, the main instance to process is the 50-node instance available in the QOPTLib library\footnote{\url{https://data.mendeley.com/datasets/h32z9kcz3s/1}} \cite{osaba2023qoptlib}.

To conduct the experimentation, a pool of similar instances, named \texttt{MaxCut\_50\_X}, is generated by eliminating a set of randomly selected \texttt{X}\% edges from \texttt{MaxCut\_50}. Next, all \texttt{MaxCut\_50\_X} instances are solved 10 times through forward annealing using D-Wave’s \texttt{Advantage\_System6.4} as a quantum device. Eventually, RA is executed, taking the best solution of each \texttt{MaxCut\_50\_X} as the classical input to solve the \texttt{MaxCut\_50} instance. 

{\color{black} To further strengthen the scope of this study, we use this initial use case to demonstrate the potentially adverse effects of counterproductive ToK on performance. For this purpose, we introduce a completely distinct instance—unrelated to \texttt{MaxCut\_50}—referred to as \texttt{MaxCut\_50\_100}.}

\begin{figure}[t]
    \centering
    \includegraphics[width=\linewidth]{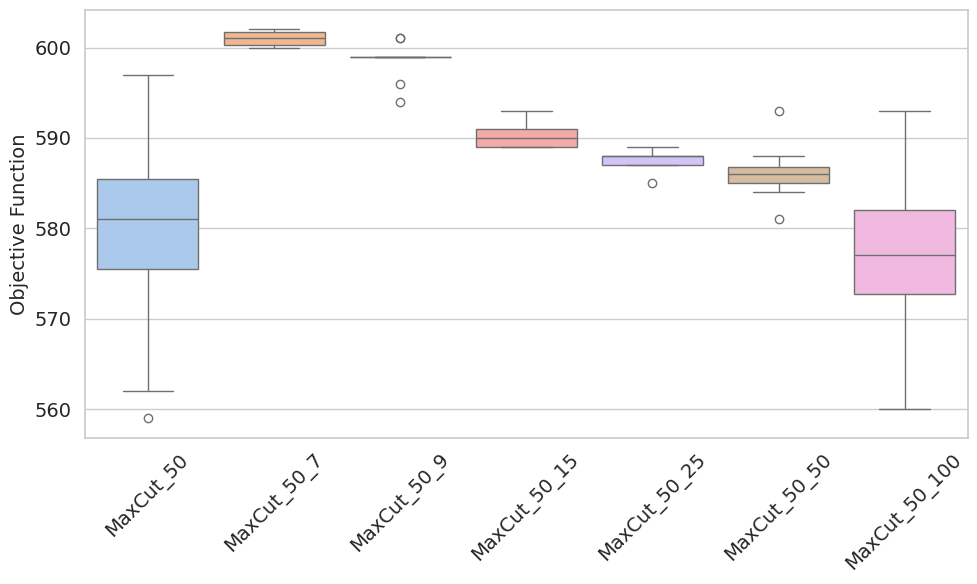}
    \caption{{\color{black}Boxplot showing the results obtained for UC1, related to the experiments on Reverse Annealing. The leftmost boxplot shows the baseline results obtained via forward annealing, while the rightmost boxplot illustrates a case of negative transfer, using as input an instance entirely unrelated to \texttt{MaxCut\_50}.}}
    \label{fig:ra}
\end{figure}

In summary, in UC1 we explore how reusing the solutions (or individuals, as seen in Table \ref{Tab:Twhat2}) of \texttt{MaxCut\_50\_X} through the RA paradigm can impact the resolution of \texttt{MaxCut\_50}. More specifically, referring to \figlabel{fig:schemes}.a, the knowledge is injected in the \texttt{annealing} step. The results of the objective function for each \texttt{MaxCut\_50}-\texttt{MaxCut\_50\_X} combination after 10 independent runs are depicted in Fig.~\ref{fig:ra} and summarized in Table~\ref{tab:exp}. To properly understand the results, it should be noted that the objective function used is the sum of the weights of the cut edges, so the bigger the value, the better the solution.

In a nutshell, analyzing both Fig. \ref{fig:ra} and Table \ref{tab:exp}, we can clearly see how the ToK among related tasks benefits the solving of \texttt{MaxCut\_50}. In fact, the resolution of the problem improves even when using the instance that is the most different in terms of composition, which is \texttt{MaxCut\_50\_50}. Also, it is interesting to observe how the quality of the results improves significantly as the similarity of the problems increases. Accordingly, transferring the instance \texttt{MaxCut\_50\_7} into instance \texttt{MaxCut\_50} yields better results than any other attempt. 

{\color{black} Finally, the results for the unrelated instance \texttt{MaxCut\_50\_100} highlight the potentially harmful effects of a negative ToK, as the RA process performs even worse than the forward annealing approach.}

Finally, given this setup of experimentation and as described in Table \ref{tab:transfer_opt}, UC1 is an example of ToK between partially different tasks with $X^f_T \sim X^f_S$ and $X^*_T \neq X^*_S$ in domains that completelly overlap with $\mathcal{D}_S \sim \mathcal{D}_T$.

\textbf{UC2: QAOA, Individual-based multitasking} In the second example, we explore the influence of applying the TO principles to QAOA in a multitasking scenario. Specifically, we address MCP by simultaneously optimizing a set of $k$ graphs. Our goal is to analyze the impact of ToK on both individual and overall performance. Thus, we have conceived an approach called \textit{Transfer-QAOA}, which takes advantage of the hybrid nature of QAOA, exploiting the classical optimization layer to facilitate information exchange between graphs without compromising the performance of the quantum algorithm.

\begin{figure}[h]
    \centering
    \includegraphics[width=\linewidth]{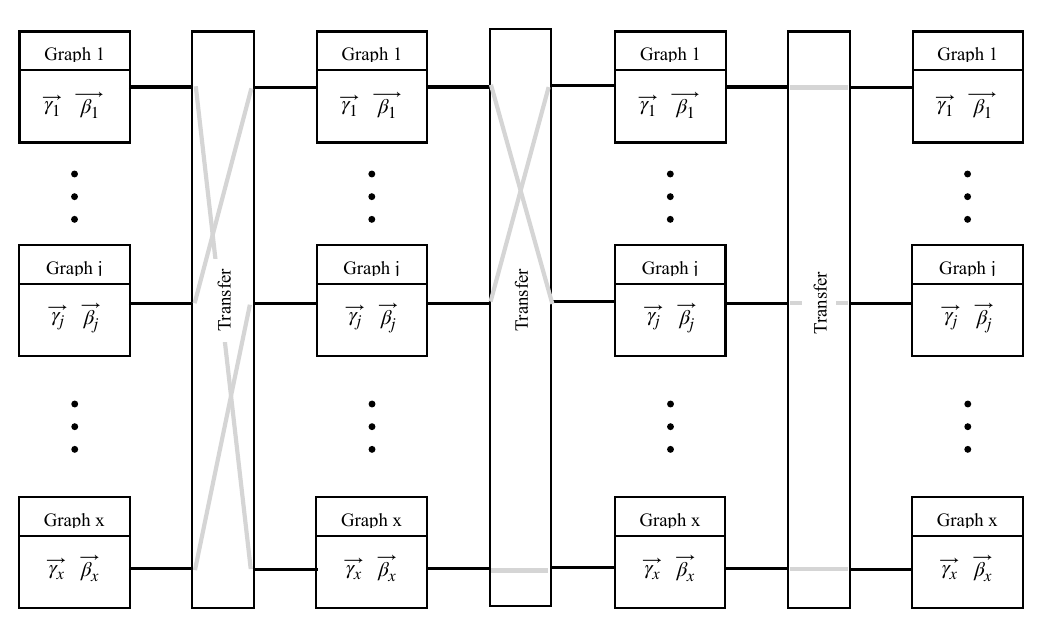}
    \caption{Schematic structure of the Transfer-QAOA algorithm. This multitasking approach leverages the optimization stage of QAOA to enable information sharing across multiple MaxCut instances. The algorithm compares the cost functions of the $k-$ instances with the parameters optimized in different cases, facilitating the transfer of information between instances when the cost function of one graph is improved by the optimization parameters of another.}
    \label{fig:transfer_qaoa_fig}
\end{figure}

\textit{Transfer-QAOA} divides the entire optimization procedure into smaller $N_T$ sub-blocks, each functioning as a standard QAOA protocol. Thus, the knowledge transfer subroutine is executed between sub-blocks (see Fig. \ref{fig:transfer_qaoa_fig}), and the information is introduced during \texttt{state manipulation} (see Fig. \ref{fig:schemes}.b). The subroutine evaluates the costs of the different $k$ graphs within the optimization parameters of the other graphs to optimize. Therefore, the parameters are shared among tasks, i.e., graphs, if they provide a smaller cost function. As for UC1, the metric to evaluate the quality of a solution is the sum of the edges cut.

Furthermore, we consider two different strategies based on the post-information-sharing process. These strategies result in the implementation of two methods. The first, \textit{transfer-static-QAOA}, proceeds with optimization using the parameters that achieve the lowest cost functions. The second, \textit{transfer-evolve-QAOA}, does not discard the previous parameters. Instead, it continues to optimize the $k$-th graph in a parallel branch. After $k'$ iterations, the costs of both branches are compared, and only the branch with the minimum cost function is retained for further optimization. This rollback mechanism prevents the optimization procedure from undergoing negative transfer.

To assess the performance of both strategies, we have applied both algorithms to a set of four MCP graphs of size 10. Each graph is coined as $MaxCut\_10\_X$, with X being an identifier to distinguish the different datasets. For the creation of these instances, the existence of overlap in the search spaces of the tasks has been sought. For this purpose, a common root graph has been used, modifying the 20\% of its edges for the creation of each instance. 

\begin{figure}[h]
    \centering
    \includegraphics[width=0.95\linewidth]{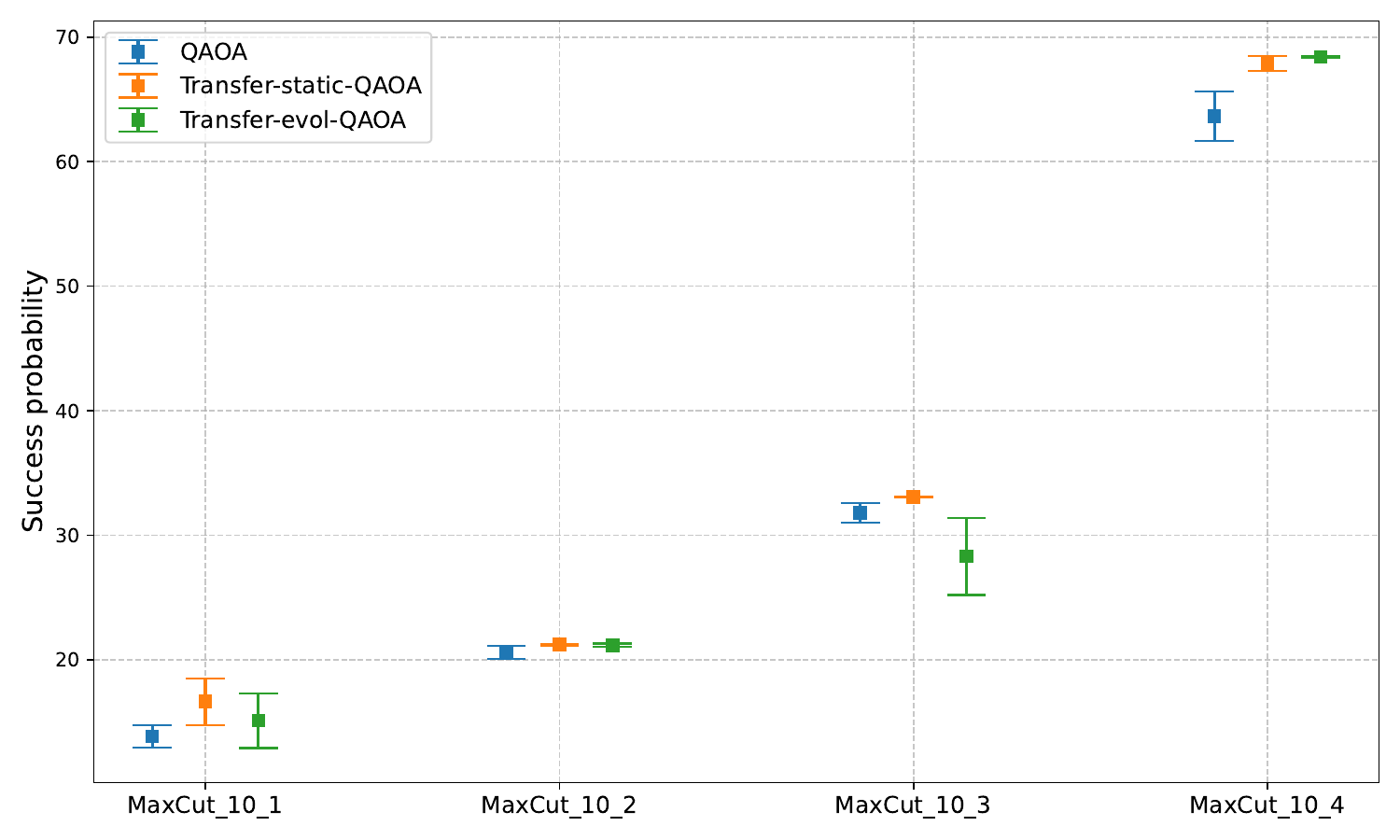}
    \caption{Success probability for the different MaxCut instances. The figure compares the statistical values and their errors for the various multitasking approaches proposed in this article, highlighting the benefits of the transfer protocol in the QAOA algorithm, particularly in the transfer-static-QAOA.}
    \label{fig:transfer_qaoa}
\end{figure}

The results obtained by \textit{transfer-static-QAOA} and \textit{transfer-evolve-QAOA} are compared with solving the graphs without any ToK. As depicted in Fig. \ref{fig:transfer_qaoa}, both approaches exhibit notable improvements in the success rate across all test cases, being the \textit{transfer-static-QAOA} more consistent with the success rate improvements. These results are based on 10 independent runs. Furthermore, as can be seen in Table \ref{tab:exp}, the overall probability of success is also improved, being the \textit{transfer-static-QAOA} the best approach.

Finally, UC2 is a case of ToK between partially different tasks with $X^f_T \sim X^f_S$ and $X^*_T \neq X^*_S$ in {\color{black} domains that partially overlap $\mathcal{D_T}\sim\mathcal{D_S}$ and $D_T \neq D_S$.}

\textbf{UC3: VQE, Parameter-based, Sequential Transfer.} In the last of the practical examples, we study the impact of transferring previously obtained knowledge on computing the ground state energy of the H$_2$ molecule. This process is carried out sequentially for different bond lengths between two hydrogen atoms, with $r \in [0.2, 2.85],\angs$ in steps of $\delta r=\SI{0.05}{\angs}$. To this end, we apply a VQE, where the \texttt{State Initialization} stage (see \figlabel{fig:schemes}.c) for a new instance is fed with the optimal, or best-found, ansatz parameters of the previous instance as the initial guess. Thus, the instance at a bond length $r+\delta r$ is initialized with the optimal parameters returned at a bond length $r$.

We use the Unitary Coupled Cluster (UCC) ansatz and Hartree-Fock mean-field for the hydrogen molecule as initial state \cite{omalley2016scalable} (see \figlabel{fig:ansatz_h2}); and SPSA~\cite{spall1992spsa} as classical optimizer using $100$ iterations. So, for the case of the H$_2$ molecule, the ansatz is given by the unitary operator $U(\theta)=e^{-i\theta X_0Y_1}$ with the initial state $\ket{\Psi_0}=\ket{01}$, therefore only one parameter $\theta$ has to be optimized. In~\figlabel{fig:theoretical_h2} we can see the theoretical optimal parameters for the H$_2$ ground state as a function of the bond length between the two atoms. Its continuity suggests that our approach should work properly.
\begin{figure}[!b]
 \centering
 \includegraphics{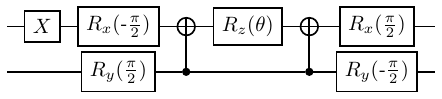}%
\caption{Hartree-Fock mean-field state initialization and UCC ansatz for H$_2$ molecule following \cite{omalley2016scalable}.}\label{fig:ansatz_h2}
\end{figure}%
\begin{figure}[!htbp]
 \centering
 \includegraphics[width=\linewidth]{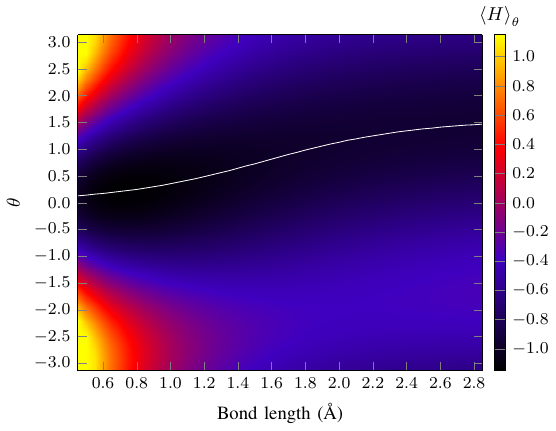}%
\caption{Energies (in Ha) of H$_2$ molecule as a function of $\theta$ and the bond length between both hydrogen atoms. The white curve denotes the theoretical ground state energy, thus the expected optimal values for $\theta$ at each length.}\label{fig:theoretical_h2}
\end{figure}

We use Qiskit \cite{qiskit} on the \texttt{ibmq\_qasm\_simulator} to simulate an ideal quantum computer, where the obtained results are compared with their corresponding exact values extracted using exact diagonalization. Results obtained through this experimentation are shown in \figlabel{fig:h2}. Within the field of computational chemistry, it is common to accept or reject a chemical prediction if its corresponding error lies below approximately $\SI{1}{kcal/mol} = \SI{0.0434}{eV} = \SI{0.0016}{Ha}$. For this purpose, we define the \emph{Iteration for Chemical Accuracy} metric (ICA) to study the convergence. The ICA returns the iteration where the energy differs from the exact expected value by less than the chemical accuracy. Therefore, the lower the ICA, the faster the convergence and thus the better the performance.
\begin{figure}[!tb]
    \centering
    \noindent\subfloat[]{\includegraphics{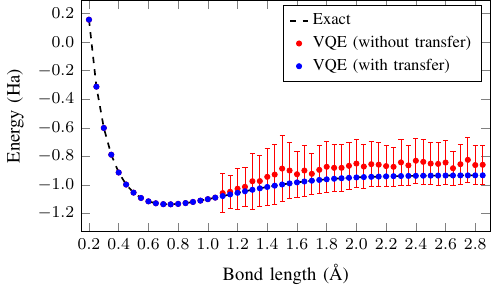}}\\%
    \noindent\subfloat[]{\includegraphics{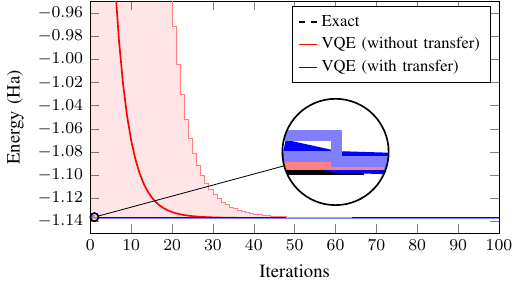}}%
    \caption{\textbf{(a)} H$_2$ molecule ground state energy, with its corresponding uncertainty, as a function of the bond length with and without transferring optimal parameters after 100 experiments. \textbf{(b)} Study of ground state energy convergence for $\SI{0.75}{\angs}$ as bond length with and without transfer of the instance optimal parameters at $r=\SI{0.70}{\angs}$ after 100 runs, where shaded area illustrates the region between the maximum and minimum values were returned and solid lines show the average results.}\label{fig:h2}
\end{figure}

In this case, and following~\tablabel{tab:transfer_opt}, UC3 is an example of transfer of knowledge across partially different tasks with $X^f_T \sim X^f_S$ and $X^*_T \neq X^*_S$ and complete overlapping domains $\mathcal{D}_S \sim \mathcal{D}_T$.

\subsection{Preliminary Insights and Reflections}

\textcolor{black}{The results obtained from the three UCs presented in the previous subsection provide consistent evidence that ToK offers measurable advantages across different quantum computing contexts. In UC1, we observe how incorporating prior knowledge of a similar instance into the annealing process enhances its performance. The tests conducted demonstrate that transfer effectiveness depends on the similarity between source and target instances, with higher similarity leading to improved solution quality and reduced variability in the solver’s output. The second use case demonstrated the utility of ToK in a multitasking setting using the QAOA. Here, knowledge transfer improved the overall success rate, illustrating how ToK can support more efficient problem-solving across parallel instances. Finally, in UC3, we observed that reusing previously optimized parameters in the VQE led to faster convergence, thereby improving the algorithm’s computational efficiency. }

\textcolor{black}{These results, while promising, are naturally constrained by the current limitations of quantum hardware. Specifically, the extent of transferable knowledge between instances is constrained by factors such as system size, the number of problems that can be parallely solved (in multitasking settings), and the availability and processability of prior knowledge. Nevertheless, the observed benefits of the proposed transfer protocols are tangible. Importantly, as ToK theory suggests, the efficacy of transfer improves with larger problem sizes and more diverse instance populations. Thus, it is reasonable to expect that the advantages of ToK will become increasingly significant as quantum hardware continues to advance.}

\textcolor{black}{In conclusion, several key insights emerge from the presented use cases. First, knowledge sharing can yield benefits across both major quantum computing paradigms—quantum annealing and gate-based models. Second, ToK can be effectively implemented in both sequential and multitasking schemes, enabling the injection of knowledge at different stages of the computational process. Finally, the nature of the knowledge being transferred is highly flexible, ranging from low-level components such as circuit parameters to higher-level abstractions like complete solutions.
}
\section{Challenges and Opportunities of Transfer of Knowledge in QC}\label{sec:challenges}

As previously discussed in this work, ToK is a mechanism to reduce the complexity and the cost of training classical ML models or solving optimization problems from scratch. %In the NISQ era, where quantum processors have a limited number of qubits and are still prone to errors and noise, the exploration of ToK schemes can be advantageous for developing more robust quantum algorithms. %
Based on the main principles outlined in \seclabel{sotaTOK}, we present several scenarios where ToK could be beneficial in QC.

\subsection{When to Transfer}

The question of \textit{when} to transfer in QC parallels that in classical computing: when it provides significant advantages in terms of algorithm performance, or when it reduces the need for extensive computation in tasks like the training of ML algorithms. However, it is important to note that QC algorithms involve different steps --depending on the paradigm--, such as state preparation, ansatz preparation and initialization, and optimization, where we can asses the potential usefulness of ToK. Moreover, for mid- or large-sized problems, extra tasks are devoted to circumventing the limited capacity of current QPUs by decomposing big problems into smaller sub-problems.

\subsection{How and Point of transfer}

When considering \textit{what} to transfer and \textit{how}, the answer becomes more extensive. However, we can again leverage techniques from classical TL and TO and adapt them to quantum computing. Regarding the pipeline of quantum algorithms mentioned above, the following are some interesting lines of investigation to pursue -- in addition to the ideas conducted in Section \ref{sec:sotaQTOK}.

\begin{itemize}
    \item \textbf{Problem conceptualization:} As previously mentioned, problem decomposition can be valuable when addressing real-world use cases in current QPUs. In this context, two decomposition techniques can be studied: \textit{i)} decompose a problem into reusable pieces. Some problem formulations are, by their nature, a sum of independent components to be solved. This is the case of the Bin Packing Problem where each bin loading is, per se, a small piece of the overall problem that, once optimized, is a potential template to reuse for future bins. \textit{ii)} Decompose a problem into similar pieces. If symmetries or approximated symmetries are identified within a complex problem, the number of  run-per-subproblem may be reduced by solving one component and adapt it to the rest of the equivalents.
    \item \textbf{State initialization (or data embedding):} In classical neural networks, ToK typically involves freezing the initial layers of a pre-trained deep neural network, as these layers capture general features. Drawing from this concept, the authors in \cite{mari2020transfer} applied a pre-trained classical neural network to embed high-dimensional data into a quantum processor, followed by a variational quantum circuit to solve the target machine learning task. This method efficiently pre-processes the data by embedding a carefully selected set of informative features into the quantum processor, bypassing the need for retraining or direct access to the raw data.
    \item \textbf{Ansatz preparation and/or initialization:} In kernel-based methods such as Support Vector Machines, ToK may involve reusing learned kernel functions, effectively transferring representations from one task to another. In the context of QC, kernels are defined through quantum embedding kernels, which map classical data into a quantum feature space to measure similarities between data points. To define efficient quantum embeddings (i.e., an appropriate ansatz) for classification purposes, the authors in \cite{hubregtsen2022training} employed the kernel target alignment concept. This concept evaluates how well a kernel function captures the true relationships in the data by comparing it to an ideal target kernel that reflects the desired classification or prediction. Although this kernel definition is highly effective, training the parameterized circuit is resource-intensive. Therefore, transferring the embedded kernel from a source task can significantly reduce the training complexity of the target alignment.
    \item \textbf{Accelerating the optimization process:} Accelerating the optimization process of hybrid algorithms can be incredibly beneficial. Depending on the optimizer used, another way to refine the optimization step is by decomposing the search space into subpopulations, thus speeding up the search process. It is important to note that VQAs can suffer from barren plateaus, a phenomenon where the gradient of the loss function diminishes exponentially as the size of the quantum system increases. This results in training stagnation because the gradients become too small for the optimizer to effectively update the parameters, leading to slow or no progress. TL can be a useful approach to mitigate barren plateaus in QC. Instead of starting with a large quantum system prone to barren plateaus, one could pre-train the quantum circuit on smaller systems or simpler tasks. The parameters from this pre-trained circuit can then be used as a starting point for the larger system, potentially avoiding the flat regions (barren plateaus) encountered with random initialization. Similarly, one can train a quantum circuit layer by layer. Parameters are optimized for the first layer and then frozen as subsequent layers are added and trained. This approach can prevent gradients from vanishing early in the process, providing a structured optimization path.
\end{itemize}

Lastly, the challenges associated with deciding \textit{how} to transfer are entirely dependent on the frequency with which the problems to be addressed arise. Although the multitasking approach can offer advantages, as seen in UC2, it is complex to devise efficient schemes that maintain the quantum properties of the qubits. In other words, sequential transfer will be advisable as long as the tasks to be addressed do not appear simultaneously.
 
As can be seen, ToK can be a promising strategy to address the challenges regarding ML and optimization algorithms in the NISQ era. By leveraging previous results and well-established techniques, it is possible to enhance the efficiency and effectiveness of quantum schemes. This approach not only accelerates the optimization process but also reduces the computational burden on quantum processors, making it a valuable tool for advancing quantum computing technologies.

\section{\color{black}{Concluding Remarks}} \label{sec:conclusions}

QC is a promising computing paradigm expected to revolutionize AI. To meet all expectations, QC will eventually need to prove advantageous in real-world settings. Fortunately, classical AI has come a long way and provides us with many lessons learned. As such, ToK has made a mark by lightening computations in scenarios where problems are repetitive or bear significant similarity. 

{\color{black} Within this context, the objectives of this work were fourfold. First, to unify the diverse concepts scattered across the field, we proposed a unified mathematical notation for Transfer Learning (TL) and Transfer Optimization (TO), establishing a common language for describing transfer processes (\textbf{O1}). Building on this foundation, we developed a structured mapping of knowledge transfer models, centered on TL and TO, and designed to address the fundamental questions of \textit{when to transfer}, \textit{what to transfer}, and \textit{how to transfer} (\textbf{O2}). %Third, we extend this theoretical foundation to the quantum computing domain, identifying key schemes where ToK principles can be applied, highlighting opportunities for cross-domain integration (\textbf{O3}).
Third, to demonstrate the practical relevance of ToK in quantum contexts, we designed and implemented three use cases across both annealing-based and gate-based paradigms (\textbf{O3}). Finally, we present a set of ML and Optimization scenarios organized around three key questions: \textit{what} knowledge should be transferred, \textit{when} in the quantum-algorithm pipeline the transfer should occur, and how to carry it out effectively (\textbf{O4}).

This work successfully addressed its four objectives through a combination of systematic literature analysis and targeted experimentation. \textbf{O1} and \textbf{O2} were achieved through a comprehensive review of classical ToK literature, which informed the development of the unified notation and mapping. %\textbf{O3} extended this framework to the quantum computing domain, identifying key schemes where knowledge transfer principles are applicable.
To fulfill \textbf{O3}, we implemented three use cases, spanning both annealing-based and gate-based quantum paradigms, that demonstrated the practical relevance of ToK. The experiments confirmed that ToK can enhance performance across diverse quantum contexts: improving solution quality and consistency in quantum annealing (UC1), increasing success rates in multitasking QAOA (UC2), and accelerating convergence in VQE through parameter reuse (UC3). Finally, \textbf{O4} is reached by providing a conceptual foundation for guiding future research and practical implementations of ToK in quantum machine learning and optimization. 

While current NISQ quantum hardware imposes limitations, the observed benefits of ToK are tangible and align with theoretical expectations. This work highlights how ToK can enhance quantum algorithm performance through strategies including problem decomposition, classical-assisted state initialization, kernel reuse for ansatz design, and structured optimization to mitigate barren plateaus. We also show that, although multitask transfer remains constrained by coherence challenges, sequential transfer proves to be a more practical and effective approach. As quantum technologies continue to evolve, these advantages are expected to scale, particularly for larger and more complex problem instances. Collectively, these findings position ToK as a promising and scalable strategy for improving the efficiency and applicability of quantum computing in the NISQ era and beyond.

However, it is important to acknowledge that the development of efficient quantum algorithms—whether based on transfer techniques or not—remains fundamentally tied to the evolution of quantum hardware. Only as quantum systems mature and become capable of solving practically relevant problems will the full potential of these approaches be realized. As such, the advantages of ToK are expected to scale in tandem with hardware progress, particularly for larger and more complex problem instances, positioning it as a promising strategy for the utility era of quantum computing.

Building on the contributions presented in this work, we aim to inspire broader adoption and ongoing refinement of ToK techniques within the quantum research community. In a landscape where hybrid approaches are not just advantageous but essential, the ability to transfer knowledge across computational paradigms and among interdisciplinary teams will be increasingly vital. As quantum computing continues to evolve, sustained collaboration and a shared conceptual framework will be key to driving innovation and unlocking its full potential.

%Building on this perspective, ToK stands out as a promising strategy to address current limitations and prepare for the opportunities of the utility era of quantum computing. By enabling more efficient algorithm design through techniques such as problem decomposition, informed state initialization, ansatz reuse, and structured optimization, ToK can significantly reduce computational overhead and enhance scalability. While its full potential is contingent on hardware advancements, the foundational benefits demonstrated in this work suggest that ToK will play a critical role in shaping future quantum applications. We hope that the contributions presented here will encourage broader adoption and continued refinement of ToK methodologies. In a landscape where hybrid approaches are increasingly essential, the ability to transfer knowledge across computational paradigms and among interdisciplinary teams will be vital. As quantum computing evolves, sustained collaboration and a shared conceptual framework will be key to driving innovation and unlocking its full potential.

}

% Within this context, the objectives of this work have been twofold:
% \begin{itemize} 
%    \item {\color{blue} \textit{To gather and present the core theories and principles} behind ToK in a brief, yet exhaustive, description and classification.} To unify the main concepts scattered throughout the vast bibliography related to this discipline, a homogenized mathematical notation for TO and TL has been proposed. This theoretical framework is the cornerstone for identifying opportunities to enhance the performance of hybrid solvers and increase the probability of success. 
%    \item \textit{To encourage quantum researchers to consider and adopt}, if applicable, ToK techniques by providing an overview of the most remarkable efforts made in classical computing and a practical demonstration of the advantages that could be gained in the field. 
%\end{itemize}
% Last but not least, and ascending to holistic thinking, knowledge transfer between members of multidisciplinary teams will also catalyze enormous advances in the QC field, where hybridization is now and will be in the future the only plausible approach.

\section*{Acknowledgments}
This work was supported by the Basque Government through Plan complementario comunicación cuántica (EXP. 2022/01341)(A/20220551). The authors thank the support of the Government of Biscay (Bizkaiko Foru Aldundia -- Diputación Foral de Bizkaia) through Lantik and its Industry Focused Quantum Ecosystem initiative, which provided access to the IBM quantum computers. We acknowledge the use of IBM Quantum services for this work. The views expressed are those of the authors and do not reflect the official policy or position of IBM or the IBM Quantum team.

\section*{Conflict of Interest Statement}

The authors declare that the research was conducted in the absence of any commercial or financial
relationships that could be construed as a potential conflict of interest.

\section*{Data availability statement}

The datasets generated for this study and the complete set of results shown are available from the corresponding author (E.V.R.) upon reasonable request. 

\bibliographystyle{IEEEtran}
\bibliography{IEEEexample}
\end{document}